\documentstyle[aps,multicol,epsf,epsfig]{revtex}

\begin{document}
\title{Recycling of quantum information:
Multiple observations of quantum clocks}
\author{
V.~Bu\v{z}ek$^{1,2}$, P.L.~Knight$^3$ and N.~Imoto$^2$
}
\address{ 
$^{1}$ Institute of Physics, Slovak Academy of Sciences,
D\'ubravsk\'a cesta 9, 842 28 Bratislava, Slovakia.\\
$^{2}$ SOKEN-DAI, The Graduate University for Advanced Studies, Shonan Village,
Hayama, Kanagawa 240-0193, Japan \\
$^{3}$ Optics Section, The Blackett Laboratory,
Imperial College, London SW7 2BW,  U.K.\\
}

\date{7 May 2000}
\maketitle
\begin{abstract}
How much information about the original state preparation 
can be extracted from a quantum system which already
has been measured? 
That is, how many independent (non-communicating)
observers can measure the quantum system sequentially 
 and give a nontrivial
estimation of the original unknown state? We investigate these questions
and we show from a simple example that quantum information is not
entirely lost as a result of the measurement-induced 
collapse of the quantum state, and
that an infinite number of independent observers 
who have no {\em prior} knowledge about the initial state 
can gain  a partial information
about the original preparation of the quantum system.
\newline
{\bf PACS numbers: 03.65.Bz, 42.50.Ar}
\end{abstract}
\pacs{03.65.Bz,42.50.Ar}

\vspace{-0.7cm}

\begin{multicols}{2}
From 
the {\em deterministic} measurement model employed in 
  classical physics
it follows that the state of the physical system is not
affected by  measurement. That is,  information about  states
of the system can be determined  with an arbitrary precision. Formally, 
from a kinematical point of view, 
this can be expressed as follows: in classical physics there are
measurements  
($m$) for which the statistics of the measurement results ($r$)
characterized by the conditional probability distribution 
$p_m(r|s)$ can be,  
for {\em all} possible states $s$ of the given classical  system,  
of the form 
\begin{eqnarray}
p_m(r|s)=\delta(s_r-s).
\label{1}
\end{eqnarray}
Moreover, these measurements do not change the state of the classical
system,  so   an arbitrary number of independent observers (i.e.
observers who do not communicate) can determine the state.

The standard Copenhagen interpretation of quantum mechanics is deeply rooted
in a model of  {\em non-deterministic} statistical  measurement \cite{Peres}.
From the kinematical point of view the quantum theory models
(predicts) 
the statistics of  results registered by   a measuring device
 when the measurement  is performed on a quantum object. 
Within  this non-deterministic model of measurement  
the conditional probability distribution  $p_m(r|s)$  can never be of the
form (\ref{1}) for arbitrary initially unknown states of a quantum system.
In  quantum mechanics the conditional probability distribution  $p_m(r|s)$
is given by the expression
\begin{eqnarray}
p_m(r|s)= {\rm Tr}\left[ \hat O_r \hat \rho_s \right], 
\label{2}
\end{eqnarray}
where the set of positive operators $\hat O_r$ which sum up to the
identity operator (i.e., the POVM) models the measuring device and the
density matrix 
characterizes the state of the quantum object being subject of the 
measurement. 

The axiomatics of quantum theory implicitly require 
that the state of the system is changed during measurement. Otherwise,
repeated measurements of the previously measured but unchanged quantum state
could reveal  yet  more  information about the state. Consequently,  
the measurement
model would eventually be equivalent
 to the standard deterministic measurement model of classical physics.
Therefore there is an additional rule which excludes
the possibility of repeated measurements.  
This additional principle is the well known 
von Neumann {\em projection} postulate.

Nevertheless it is an 
interesting question to ask how much information about the original state 
is ``left'' in the system which already has been measured. 
That is, how much information about the preparation can be extracted
from the system by a second observer who does not communicate with
the first observer. A further question we 
 would like to understand is whether from the axioms
of quantum theory we can obtain a ``classical-like'' picture when 
a physical system in an unknown state 
can be repeatedly measured, yet still retain  information about the
original state preparation.
In what follows we   analyze  a simple example which illuminates
these two questions.

First we specify the task of a measurement.
In  measurement we wish to determine 
some parameters of the state of a quantum 
system which correspond to a symmetry group.
As an  example, consider  the position measurement which is connected
with the group of translations, or measurement of the angle of
orientation connected with the group of rotations. In what follows 
we  analyze the
simplest example of a continuous
parameter $\varphi \in \langle 0,2 \pi \rangle$, the phase,
which parameterizes the
group of rotations in the two-dimensional space of the $U(1)$ group.
To make our discussion more physical we  consider a model
of  optimal quantum clocks  discussed in our previous work
\cite{Buzek}. We will analyze the situation when the observers
have no {\em a priori} knowledge about the original state preparation.
This means that the prior phase distribution is constant and equal
to $1/2\pi$

In our previous paper we studied the problem of 
building an optimal quantum clock from an ensemble of $N$ ions
\cite{Wineland}. 
In particular we
assumed an ion trap with $N$ two-level ions
all in the ground state 
$|\Psi\rangle=|0\rangle\otimes\dots \otimes |0\rangle$.
This state is an eigenstate of the free Hamiltonian and thus cannot record
time (phase). Therefore the first step in building a clock was to bring the
system to an appropriate initial (reference) 
state $\hat\Omega$ which is not an energy
eigenstate. For instance, one can apply a Ramsey pulse whose shape and
duration is chosen such that it puts all the ions in the product state
\begin{eqnarray}
\hat{\Omega}=\hat{\rho}^{\otimes N}, 
\label{3}
\end{eqnarray}
with $\hat{\rho}=|\psi\rangle\langle \psi|$ and $|\psi\rangle
= (|0\rangle +| 1\rangle)/\sqrt{2}$.
After this preparation stage,
the ions evolve in time according to the Hamiltonian evolution
$\hat \Omega(t) = \hat U(t) \hat \Omega \hat U^\dagger(t)$, 
$\hat U(t)=\exp\{-i t \hat{H}\}$ (we use units such that $\hbar=1$).
Therefore these ions  can be viewed as a time-recording device. 
The task is to determine this time $t$ (or equivalently the corresponding
phase) by carrying out a measurement on the ions.
Note that because of the  indeterminism of quantum mechanics
it is impossible, given a {\it single} set of $N$ two-level ions, to
determine the elapsed time with certainty. As we have shown earlier
\cite{Derka} one can find an optimal measurement (see below) with the
help of which  information about the phase can be most optimally
``extracted'' from a system of $N$ identically prepared spin-1/2 particles.
The ability of the system to retain information about the phase (time) depends
very much on the choice of the initial reference state $\hat{\Omega}$.
For instance, if this state is an eigenstate of the total Hamiltonian, 
the system is not able to record (keep) time information.
In Ref.~\cite{Buzek} we addressed   the
question of which is  the most appropriate initial state  
$\hat\Omega$ of $N$ spin-1/2 particle which ``keeps'' the record
of phase in the most reliable way. 
In other words,  what are the optimal quantum
clocks and what is the performance of such quantum clocks when compared
with  classical clocks. 

In the present paper we investigate   
another aspect of this comparison. Namely, we  discuss the ``robustness''
of quantum clocks with respect to repeated measurements performed on them.
Classical clocks, as all classical objects, do not change their 
state or behavior  when they are observed. As we stated above  
this is no more true for   quantum objects. This has consequences 
for the functioning of our proposed quantum clocks. In particular, one
can ask whether quantum clocks may be  robust enough in the sense that
repeated readout of the time, let us say by many independent and
non-communicating observers, can provide  reliable information
(if any) about the time to all of them. 
In order to find quantitative answers to our questions  
let us recall briefly  the details of how  time is read out from
our quantum clocks.

In general, a quantum-mechanical 
measurement is 
 described by a POVM \cite{Peres,Helstrom,Holevo} which is 
a set $\{\hat O_r\}_{r=1}^ R$ of positive Hermitian operators such
that $\sum_r \hat O_r = \hat{\openone}$. Because such measurement is in
general non-deterministic,
to each outcome $r$ of the measurement we associate an estimate $t_r$
of the time elapsed. The difference between the estimated time $t_r$
and the true time $t$ is quantified by a cost function $f(t_r - t)$
\cite{Holevo}.
Here we note that because of the periodicity of the clock, $f$ has to be
periodic. We also take $f(t)$ to be an even function
to ensure a non-zero average. 
Our task is to minimize the mean value of the cost function
\begin{eqnarray}
\bar f = \sum_r \int_0^{2\pi} 
{\bf Tr}[\hat O_r\ \hat \Omega(t) ] f(t_r-t) \frac{d t}{2\pi}. 
\label{4}
\end{eqnarray}
Following Ref.~\cite{Buzek}
we expand the cost function in a Fourier series:
\begin{eqnarray}
f(t) = w_0 - \sum_{k=1}^\infty w_k \cos k t .
\label{5}
\end{eqnarray}
The essential hypothesis made by Holevo \cite{Holevo}
is the positivity of the Fourier
coefficients: $w_k \geq 0$, $(k=1,2,...) $
Without loss of  generality (see Ref.~\cite{Buzek}) we can assume 
the initial (reference) state $\hat\Omega =
|\psi\rangle\langle\psi|$, $|\psi\rangle = 
\sum_m a_m |m\rangle$  to be a pure state (in what follows 
we make a phase convention such that $a_m$ are real and positive). 
In this case for the mean cost (\ref{4}) we find
the bound \cite{Holevo}
\begin{eqnarray}
\bar f \geq w_0 -{1 \over 2} \sum_{k=1}^\infty w_k \sum_{m,m' \atop
|m-m'|=k}
a_m a_{m'}. 
\label{6}
\end{eqnarray}
In this last expression,  
equality is attained only if the measurement is of the form
\begin{eqnarray}
\hat O_r = p_r |\Psi_r\rangle\langle\Psi_r|\quad ; 
\label{7}
\end{eqnarray}
with 
$p_r \geq 0$ and $\sum_{r} \hat O_r = \hat{\openone}$, where  
\begin{eqnarray}
|\Psi_r\rangle = e^{ i t_r  \hat H} |\Psi_0\rangle ,\qquad
|\Psi_0\rangle = {1 \over \sqrt{N+1} } \sum_{m=0}^N |m\rangle,
\label{8}
\end{eqnarray}

Holevo \cite{Holevo}  originally 
considered covariant measurements in which  times $t_r$
take a continuum of values between $0$ and $2 \pi$. But as shown in
Ref.~\cite{Derka} the completeness relation can also be satisfied by taking
a discrete set of times $t_r = {2 \pi r \over N +1}$,
$r=0,...,N$. The states $|\Psi_r\rangle$ form an orthonormal basis of the
Hilbert space, and the corresponding measurement is therefore a von Neumann
measurement. This is important for applications, because it means that
it is not necessary to use an ancilla to make the optimal measurement.

We should also 
note that the states $|\Psi_r\rangle$ are the eigenstates of the
Pegg-Barnett Hermitian phase operator\cite{Pegg}. For this reason we
call them ``phase states''. In the basis $|m\rangle$ of the symmetric
subspace of $N$ two-level ions (spin-1/2 particles) 
they can be expressed as
\begin{eqnarray}
|\Psi_r\rangle= 
 {1 \over \sqrt{N+1} } \sum_{m=0}^N {\rm e}^{i\frac{2\pi}{N+1} r m}|m\rangle.
\label{9}
\end{eqnarray}                                           
The fact that the optimal measurement can 
be chosen as a von Neumann measurement is  important for our 
further considerations. This is due to the fact that 
the state immediately after the measurement is uniquely determined by 
the von Neumann projection postulate.

Before we proceed further we turn our attention to the fact that 
from the  positivity of $w_k$ it follows that 
 not all cost functions are covered by the above result. Specifically,
the quadratic deviation
$t^2$ cannot be used. On the other hand for small $t$
it can be well approximated by the cost function 
$4 \sin^2 {t \over 2}\simeq t^2$. Therefore, in what follows 
we will use the cost function, 
$4 \sin^2 t/2$. In this case 
$\bar f \simeq \Delta t^2$.

Once we have determined the optimal measurement 
we have to specify the initial (reference)  state
$\hat \Omega$ of our system.
As discussed in Ref.~\cite{Buzek},   by an
appropriate choice of this state one can substantially improve
the quality of  estimation. However, this concerns the
estimation performed by the first observer (see footnote~\ref{foot1}).
The subsequent observers
will actually always observe only rotated phase states.
These are generated in the von Neumann measurement performed
by the previous observer and subsequent time evolution. 
Therefore, in order to simplify our calculations 
we will assume that  the initial (reference) state $\hat \Omega$
is the phase state $\hat \Omega=|\Psi_0\rangle\langle \Psi_0|$
given by Eq.~(\ref{8}).

Let us  study now how the independent observers measure a system
of $N$ spin-1/2 particles initially prepared in an unknown state
obtained by the rotation of the reference state (\ref{8}). 
 As far as the first is 
observer concerned, the problem has been  already solved (see 
\cite{Derka}) and the mean cost
Eq.~(\ref{4}) can be calculated. For our  reference state
$\hat \Omega=|\Psi_0\rangle\langle \Psi_0|$ the mean cost as a function
of number $N$ of spin-1/2 particles  is given by the expression
\begin{eqnarray}
\Delta t^2 \simeq \bar{f}(N;1)= 2\left[1-\frac{N}{N+1}\right]=\frac{2}{N+1}.
\label{10}
\end{eqnarray}
We see that the mean cost when a single measurement is performed ($N=1$)
takes the  value $\bar{f}(N=1;1)=1$. On the contrary, for
$N\rightarrow \infty$ the mean cost is equal to zero. Specifically,  
for large $N$ the variance $\Delta t$ goes to zero as 
$1/\sqrt{N}$. This is far of being optimal\footnote{
As shown in Ref.~\cite{Buzek} in order to  make this variance minimal
we should take the reference state to be 
\begin{eqnarray}
|\Psi_{opt}\rangle \simeq {\sqrt{2} \over \sqrt{N+1}}
\sum_{m=0}^N \sin{ \pi (m+1/2) \over N+1}
|m\rangle.
\label{11}
\end{eqnarray}
In this case the
cost decreases for large $N$ as
$\bar f_{opt} \simeq {\pi^2 \over ( N+ 1)^2} $
corresponding to
$\Delta t_{opt} \simeq {\pi \over ( N+ 1)}$. 
\label{foot1}
}. 
Nevertheless, as  our task is to study how much information
subsequent observers can gain we are not over-worried  about the
optimality of the of the preparation of the reference state. 
Our further result can be understood as a lower bound 
and the optimization can be performed rather straightforwardly anyway.

Now we turn our attention to subsequent observers. 
We have assumed our  observers
do not communicate. If they do then the first observer can broadcast
the result of his measurement (or, which is equivalent he can broadcast
the orientation of his apparatus) 
 and there is no need for subsequent observers
to perform any measurement,  because they know that they cannot perform better
than this  first observer. 
To describe the mean cost of the estimation of subsequent observers, we
have to modify 
in  Eq.~(\ref{4})
the conditional probability distribution 
$p_1(r|t)={\bf Tr}[\hat O_r\ \hat \Omega(t) ]$ characterizing the 
measurement statistics of the  observer.
This is because the $(k+1)$-st observer does not observe
the original state $\hat \Omega(t)$. He can only measure the 
state generated via the measurement performed by the previous
$k$-th  observer. In addition, the following random factors
enter the game:  Firstly, the 
 $(k+1)$-st observer does not have  full
information about the choice of the measuring apparatus of the
$k$-th observer. Although all observers posses the optimal measuring
apparatus of the same construction ( corresponding to the optimal
von Neumann measurement) there is one parameter they can choose at random.
Namely,  if we take the POVM characterized by the set of projectors
$\hat O_r=|\Psi_r\rangle \langle \Psi_r |$, $r=0,\dots,N$ 
and we rotate them all by the same transformation 
$\hat U(\alpha)=\exp\{-i\alpha \hat H\}$
we get a new POVM $\hat O_r^{\alpha}=\hat U(\alpha)\hat O_r
\hat U^\dagger(\alpha)$ which also corresponds to the 
optimal measuring apparatus.
It is this information about the angle 
$\alpha '  \in \langle 0,2\pi\rangle$ characterizing  
the ``actual orientation''
of the $k$-th measuring apparatus 
which is not available to the $(k+1)$-st
observer. The second piece of information which is not available to the
$(k+1)$-st observer is, which of the possible outcomes $r'$ of the measurement
was detected by the $k$-th observer. Finally, the actual time
$t'$ when this measurement was performed is also unknown (however,
as we will see in a moment, this is not important for our consideration).
Taking into account  these random factors the required conditional
probability distribution $p_{k+1}(r|t,\alpha)$ (we have included the 
parameter $\alpha $ into the conditional probability distribution) 
reads
\begin{eqnarray}
p_{k+1}(r|t,\alpha) &=& \sum_{r'=0}^N \int_0^{2\pi}\ 
 p_k(r'|t',\alpha ')\ {d\alpha ' \over 2\pi} 
\nonumber 
\\
& \times & {\bf Tr} \left[ \hat O_r^{\alpha } \  
\hat U(t-t') \hat O_{r'}^{\alpha '} \hat U^\dagger(t-t')\right]. 
\label{12}
\end{eqnarray}
It is easily seen that this can be simplified as
\begin{eqnarray}
p_{k+1}(r|t, \alpha)= {\bf Tr} \left[\hat \Omega_{k+1}(t) \  \hat O_r^\alpha 
\right],
\label{13}
\end{eqnarray}
where 
\begin{eqnarray}
\hat \Omega_{k+1}(t) = \sum_{r'=0}^N \int_0^{2\pi} p_k(r'|t,\alpha ') \ 
\hat O_{r'}^{\alpha '}\ {d\alpha ' \over 2\pi}.
\label{14}
\end{eqnarray}
The last transformation is possible because 
$p_k(r'|t',\alpha ')=p_k(r'|t,\alpha '+t-t')$
and the integration with respect to $\alpha '$ ensures 
 that the shift 
$(t-t')$ is irrelevant.

Using the iterative definition given by Eqs.~(\ref{13}) and (\ref{14})
together with the definition  for the mean cost~(\ref{4}) we 
calculate the precision of the measurement of  time 
performed with  the quantum
clocks as a function  of the number of qubits $N$ 
and the number of subsequent observers  $k$:
\begin{eqnarray}
\Delta t^2\simeq \bar{f}(N;k)=2\left[1 - \left({N \over N+1}\right)^k\right].
\label{15}
\end{eqnarray}
This is the main result of our paper. 
We stress that the above result holds for the reference state
 corresponding
to the phase state (\ref{8}) and the case that observers have no {\em 
a priori} knowledge about the initial state preparation.
It can be generalized to the case
when the initial reference state is take to be the optimal state
(\ref{11}) - unfortunately in this case we are not able to find
a solution in an elegant closed analytical form.

Let us summarize our result: We have shown that quantum information
can be recycled in a sense, that by performing a measurement on
quantum systems which have already been measured independent 
observers can still
obtain non-trivial information about the original preparation 
of the quantum system (i.e. the quantum information). The larger is 
the ensemble ( $N$) the   more robust is the quantum system with
respect to subsequent 
measurements. Obviously, as  follows from Eq.~(\ref{15}) 
for the $(k+1)$-th observer the mean cost 
of the estimation will be larger than for the $k$-th observer. 
From the point of view
of information stored in the system, in 
the large-$N$ limit the quantum system
behaves very classically, i.e. an infinite number of independent observers
who have no prior knowledge about the state preparation 
can precisely measure the state of the system.

%\acknowledgements
One of us (V.B.) thanks Rado Derka and Serge Massar 
for helpful discussions. 
This work was in part supported by the Royal Society, by 
the  IST project EQUIP under the contract
IST-1999-11053, by the UK Engineering and Physical Sciences Research
Council, and by the CREST, Research Team
for Interacting Career Electronics.

\vspace{-0.5cm}

\end{multicols}


\begin{references}
\vspace{-1.5cm}

\bibitem{Peres}
A. Peres {\it Quantum Theory: Concepts and Methods}
(Kluwer, Dordrecht, 1993).

\bibitem{Buzek}
V. Bu\v{z}ek, R. Derka, and S. Massar, 
{\em Phys. Rev. Lett.} {\em 82}, 2207 (1999).
We do not address here questions concerning the improvement
of frequency standards obtained through the use of entanglement
as discussed by S.F.Huelga et al. 
{\it Phys. Rev. Lett.} {\bf 79}, 3865 (1997).



\bibitem{Wineland}
D.J. Wineland, C. Monroe, W.M. Itano, D. Leibfried, B.E. King, and 
D.M. Meekhof, 
{\it J. Res. NIST} {\bf 103}, 259 (1998).


\bibitem{Derka}
R. Derka, V. Bu\v zek, and A.K. Ekert, { Phys. Rev. Lett} {\bf 80},
1571 (1998).


\bibitem{Helstrom}
C.W. Helstrom, {\it Quantum Detection and Estimation Theory}
(Academic Press, New York, 1976).


\bibitem{Holevo}
A.S. Holevo, {\it Probabilistic and Statistical
Aspects of Quantum Theory} (North-Holland, Amsterdam, 1982).



\bibitem{Pegg}
D.T. Pegg and S.M. Barnett, { Europhys. Lett.} {\bf 6}, 483 (1988).


\end{references}
\end{document}